\documentstyle[12pt,camready,epsfig]{article}
\title{NEUTRON CAPTURE AND NEUTRON HALOS}
\author{A. MENGONI \footnote{\noindent
Permanent address: ENEA, Applied Physics Section, v. Don Fiammelli 2,
I-40128 Bologna, Italy. e-mail: {\tt mengoni@rikvax.riken.go.jp}}}
\address{Radiation Laboratory, 
The Institute of Physical and Chemical Research (RIKEN),\\
2-1 Hirosawa, Wako, Saitama 351-01, Japan}
\author{T. OTSUKA, T. NAKAMURA and M. ISHIHARA}
\address{Department of Physics, The University of Tokyo, \\
7-3-1 Hongo, Bunkyo-ku, Tokyo 113, Japan}
\begin{document}
\maketitle
\abstracts{\noindent{\bf{Abstract}}: 
The connection between the neutron halo observed in light neutron
rich nuclei and the neutron radiative capture process is outlined. 
We show how nuclear structure information such as spectroscopic  
factors and external components of the radial wave
function of loosely bound states can be derived from the neutron
capture cross section. The link between the direct radiative capture 
and the Coulomb dissociation process is elucidated.}

\section{Introduction}
The neutron capture process in a light neutron-rich  nucleus  may
be  viewed  as  a creation of a halo state whenever the final
bound orbit into which the neutron is captured has a strong
single-particle {\it s}-wave component and a small binding energy.  
It is then possible to investigate the properties of the halo state 
by analyzing the $(n,\gamma)$ process which leads to its formation.  
There are certain conditions which must be satisfied 
to apply this technique.
For  example,  the  neutron capture must proceed through a direct
transition and should not be mediated by the formation of a  compound  
state. Such a process is referred to as the direct radiative 
capture process (DRC). 
In this situation, the capture mechanism can be  described by a theoretical
model  in  such a way that the structure information derived does
not depend on the capture mechanism.

As an example, we will show in detail in the next paragraphs that
the  neutron  capture cross section as a function of the incident
neutron kinetic energy can  be  considered  as  a  Fourier-Bessel
transform of the radial wave function.  By  inverting  the
$(n,\gamma)$ cross section it is therefore possible to derive  the
radial component of the halo state wave function.

In addition to nuclear structure investigations, the modeling
of the neutron capture process finds major applications 
in nuclear astrophysics. In fact, several crucial
$(n,\gamma)$  reaction rates are required, for example, 
in the network calculation of the nucleosynthetic processes 
in the inhomogeneous big-bang scenario \cite{Ka95}.  
Some of these reaction rates may be derived from
laboratory experiments but some others  need  to  be  supplied  by
theoretical  calculation. In any case, the theoretical models are
necessary to complement the experimentally derivable quantities
to obtain the reaction rates in the temperature range of
interest for astrophysical applications.

\section{Direct radiative capture (DRC)}
A minimal description of the direct 
radiative capture model (DRC) is given here.
Details and references to alternative approaches
can be found elsewhere \cite{Otx94}${}^{-}\,$\cite{Krax96}.
The capture cross section for emission of
electric dipole radiation (E1) in the
transition from a state in the
continuum ({\it c}) to one of the
bound states of the residual nucleus ({\it b})
is given by 
\begin{equation}
\sigma_{n,\gamma} = \frac{16 \pi}{9 \hbar v}
k_{\gamma}^{3} \bar{e}^2
\vert Q^{(E1)}_{b \leftarrow c} \vert^{2}
\label{capture}
\end{equation}
where $v$ is the neutron-nucleus relative velocity,
$k_{\gamma} = \epsilon_{\gamma}/\hbar c$ 
the emitted $\gamma$-ray wave number
and $\bar{e}$ the neutron effective charge.
The calculation of the matrix elements 
\begin{equation}
{{\it Q^{(E1)}_{b \rightleftharpoons c}}} =
< \Psi _{c} \vert \hat{T}^{E1} \vert \Psi _{b} >
\label{ME1}
\end{equation}
where $\Psi_{b}$ is the bound-state
wave function and $\Psi_{c}$ the wave function
for the neutron in the continuum, is not an easy task, 
in general.
They can, however, be readily evaluated under
the particular condition in which the
bound state has a strong single particle
component. In this case, they can be decomposed into
the products of three factors
\begin{equation}
Q^{(E1)}_{b \rightleftharpoons c} \equiv
\sqrt{S_b} \ A_{b,c} \ {\cal I}_{b,c} 
\label{ME2}
\end{equation}
where $S_b$ is the spectroscopic factor of the bound state,
$A_{b,c}$ is a factor containing only angular momentum and
spin coupling coefficients and the radial overlap
\begin{equation}
{\cal I}_{l_{b}l_{c}} = \int\limits_{0}^{\infty}
u_{l_b}(r) r w_{l_c}(r) dr
\label{eqi1}
\end{equation}
can be evaluated using some potential model for the
calculation of the radial wave functions $u_{l_b}(r)$
and $w_{l_c}(r)$.

We have recently applied the DRC model to the 
calculation neutron capture cross sections of several 
light nuclei (${}^{12}$C, ${}^{13}$C, ${}^{14}$C, 
${}^{16}$O, ${}^{10}$Be). 
The results have been reported elsewhere 
\cite{Mex96a}${}^{-}\,$\cite{Mex96c}
and the general conclusion is that the DRC model can 
well describe the neutron capture process in the
energy region from thermal up to several hundred
of keV. Other calculations made using
an essentially equivalent treatment have been
recently performed \cite{Krax96} with results 
quite similar to those obtained by our group.

\section{Neutron capture and Coulomb dissociation}
The matrix elements $ Q^{(E1)}_{b \rightleftharpoons c} $
have been written with a double arrow because they are the
same matrix elements which can be obtained
in the inverse of the $(n,\gamma)$ reaction: 
the Coulomb dissociation. 
In a Coulomb dissociation experiment, 
a break-up process of the nuclei in the incident beam
is induced by the Coulomb field generated 
by a high-{\it Z} target. 
Under defined kinematic conditions,
the $B(E1)$ strength distribution for the
dissociation of the incident
${}^{A+1}$X nucleus into ${}^{A}$X$ + n $
is measured. The $B(E1)$ strength distribution 
is related to the neutron capture cross section by
\begin{equation}
\frac{dB(E1)}{d E_{x}} = \frac{9}{ 16 \pi^3} \
\frac{k_{n}^{2}}{k_{\gamma}^{3}} \
\frac{ 2J_{c} + 1}{2J_{b}+1} \ \sigma_{n,\gamma}
\end{equation}
where $E_{x}$ is the excitation energy (defined as
the sum of the neutron-residual nucleus relative energy 
plus the neutron binding energy), 
$k_{n}$ is the neutron wave number in the continuum,
$J_{b}$ is the total angular momentum
of the bound state and $J_{c}$ the spin of the residual
nucleus in the continuum. The $B(E1)$ strength distribution
is therefore related to the matrix elements by
\begin{equation}
\frac{dB(E1)}{d E_{x}} = 
\frac{k_{n}^{2}}{\pi^2 \hbar v} \ \bar{e}^2 \
\frac{ 2J_{c} + 1}{2J_{b}+1} \ 
\vert Q^{(E1)}_{b \rightarrow c} \vert^2
\label{dbe1b}
\end{equation}
It must be noted here that, while in a neutron capture measurement
all the available bound states can be in principle populated by the
incident neutron, in a Coulomb dissociation experiment, only
transitions originating from the ground state of the
nuclei in the incident beam are possible. 
In turn, the great advantage of the Coulomb dissociation 
measurement is that the E1 matrix elements of radioactive nuclei
can be investigated. In fact, with the recent developments
made with radioactive beam facilities,
E1 matrix elements of unstable neutron rich nuclei up to the
drip-line have been measured. 
A first noticeable example of the application of this method 
has been the measurement of the Coulomb dissociation 
of ${}^{11}$Be \cite{Nax94}. This is a
very well known example of halo nucleus.
In fact, its ground state is bound by
only 505 keV and is dominated by the
\mbox{$ | {}^{10}\mbox{Be}(0^{+}) \otimes (2s_{1/2})_{\nu} >$}
configuration. The $ dB(E1) / d E_{x}$
strength distribution can be well reproduced \cite{Otx94,Mex96b} 
by calculation made using Eq.~\ref{dbe1b} and the
matrix elements can be used to derive information
on the ${}^{11}$Be ground-state wave function (see below).

\section{Nuclear structure information}
A noticeable property of the overlap integral 
in Eq.~\ref{eqi1} is that whenever the continuum
state wave function $\Psi_{c}$ can be well 
approximated by a two-body wave function,
detailed information on the bound state
can be derived from given overlaps.
Moreover, it can be seen that when the 
bound state wave function is a $ l = $ 0 orbit 
(and the initial state must be in this case a 
$l = $ 1 state because of the E1 selection rules), 
the overlap takes place essentially 
in the region outside the nuclear radius. 
This is simply due to the fact that
{\it p}-wave neutrons do not penetrate 
into the internal nuclear region because 
of the centrifugal barrier. 
In turn, the wave function of a bound {\it s}-orbit  
extends significantly outside the nuclear surface 
and can be well described by a Yukawa-tail of
type $ \psi_{b}(r) \sim \exp{(- \chi r)}/r $ where
\mbox{$ \chi = \sqrt{2\mu E_{b}} / \hbar $}
(here, $\mu$ is the reduced mass of the two-body
system and $E_{b}$ the energy of the bound state). 
This is the typical representation of a halo-state
wave function in the asymptotic region \cite{HanJon87}.

From the same considerations it also follows that 
the overlap is insensitive to 
the neutron-nucleus interaction in the continuum 
and the wave function $w_{l_c}(r)$ can be 
well described by the $ l = $ 1 component of a plane-wave. 
This can be verified numerically. 
%
%%%%%%%%%%%%%%%%%%%%%%%%%%%
%
\begin{figure}
\begin{center}
\leavevmode
\hbox{
\epsfxsize=10.0cm
\epsfysize=9.0cm
\epsffile{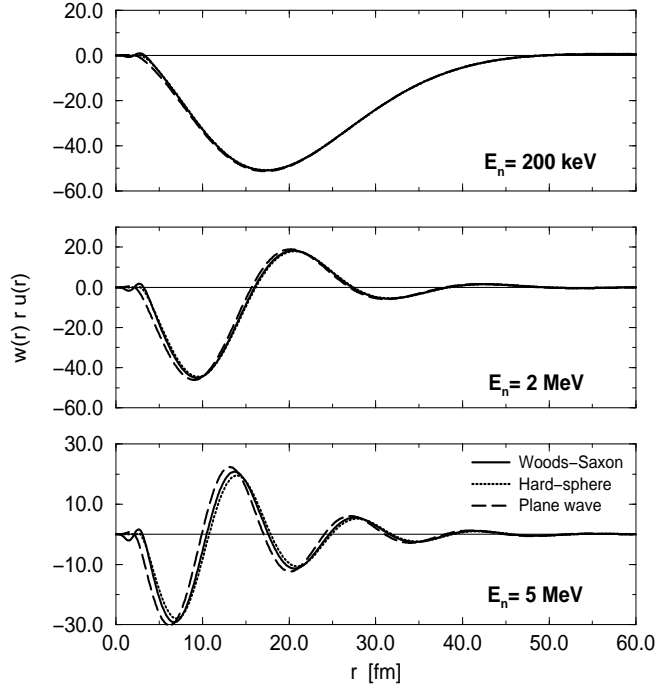}}
\end{center}
\caption{Radial part of the E1 matrix elements 
(integrand of Eq.~4). The parameters used
to calculate the wave functions are given in the text.
}
\label{fig:overlap}
\end{figure}
%
%%%%%%%%%%%%%%%%%%%%%%%%%%%
%
In figure \ref{fig:overlap}
we show the integrand of $ {\cal I}_{l_{b}l_{c}} $
for a case in which the bound state is a {\it s}-orbit
bound by 500 keV in a Woods-Saxon potential 
of radius $R = $2.83 fm, depth $V_0 = $ 57.8 MeV
and diffuseness $d = $ 0.62 fm. Three different
potentials have been considered for the
neutron-nucleus interaction in the continuum.
They represents a kind of limiting cases and
are respectively a Woods-Saxon potential with
the same parameters as those of the bound
state, a hard-sphere potential of
radius $R = $2.83 fm and a case with
no-potential (plane-wave). For neutron
energies in the continuum up to 5 MeV
there is essentially no difference 
(always less than 5\% in the specific case) 
in the overlap. 
This demonstrates and extend the validity of
our conclusion \cite{Mex95} on the independence of the
$p \rightarrow s$ E1 matrix elements 
on the neutron-nucleus
interaction in the continuum,
for neutron energies up to 5 MeV.

\subsection{Spectroscopic factor}
An important consequence of the statement proven
above is that, assuming that a reliable model
is available to calculate the bound state
wave function $ u_{l_b}(r) $, the spectroscopic
factor $S_b$ can be derived
from a measurement of the capture cross section
(see Eqs.~\ref{ME2} above).
This procedure, already established and applied
in proton capture measurements can be used in
the $(n,\gamma)$ case as well. In principle, a
single-energy experimental value would be sufficient
to derive $S_b$ from a measurement of $\sigma_{n,\gamma}$.
However, cross section values in a wide energy
range may improve the reliability of the
obtainable values. Of course a Coulomb 
dissociation measurement would be altogether
equivalent. As an example of a possible application
of this method we show here the
calculation of the E1 strength distribution
for the Coulomb dissociation of ${}^{19}$C
into ${}^{18}$C$ + n$. It can be clearly 
seen from the figure that in this case
the structure of the ground-state of ${}^{19}$C
could be deduced from a Coulomb dissociation
experiment. In fact, three possible configurations
are available for the ground-state: a 
\mbox{$ | {}^{18}\mbox{C}(0^{+}) \otimes (2s_{1/2})_{\nu} >$}
or 
\mbox{$ | {}^{18}\mbox{C}(0^{+}) \otimes (1d_{5/2})_{\nu} >$}
and
\mbox{$ | {}^{10}\mbox{C}(0^{+}) \otimes (1d_{3/2})_{\nu} >$}.
The strength distribution for the latter two configurations
differs from that of the most likely former one by more than
a factor of 40. The energy dependence is also quite
different for the different configurations. As has been
recently proposed \cite{Bax95}, the structure of the ${}^{19}$C  
could be that of a halo state and its spectroscopic
structure could be deduced from the measurement of the
$B(E1)$ strength distribution.
%
%%%%%%%%%%%%%%%%%%%%%%%%%%%
%
\begin{figure}[t]
\begin{center}
\leavevmode
\hbox{
\epsfxsize=10.0cm
\epsfysize=9.0cm
\epsffile{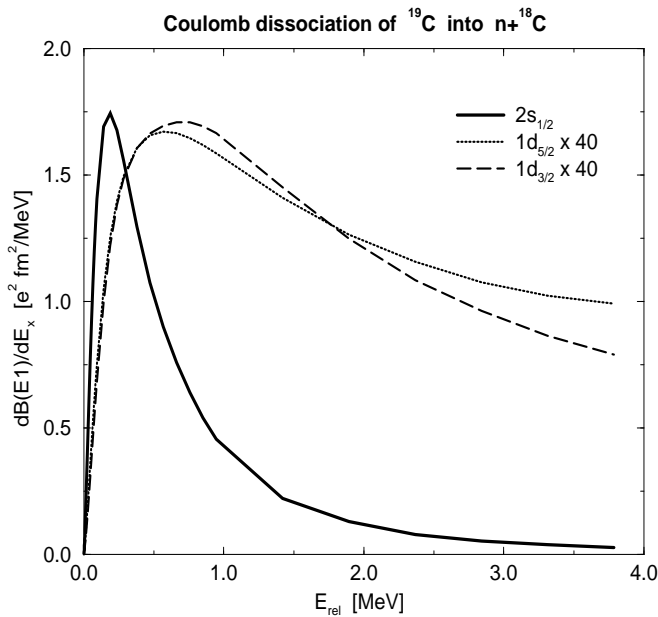}}
\end{center}
\caption{E1 strength distribution 
for the break-up of ${}^{19}$C calculated assuming
three different configurations for the ground-state.
}
\label{fig:O_potential}
\end{figure}
%
%%%%%%%%%%%%%%%%%%%%%%%%%%%
%
\subsection{Radial wave function}
\begin{figure}
\begin{center}
\leavevmode
\hbox{
\epsfxsize=10.0cm
\epsfysize=9.0cm
\epsffile{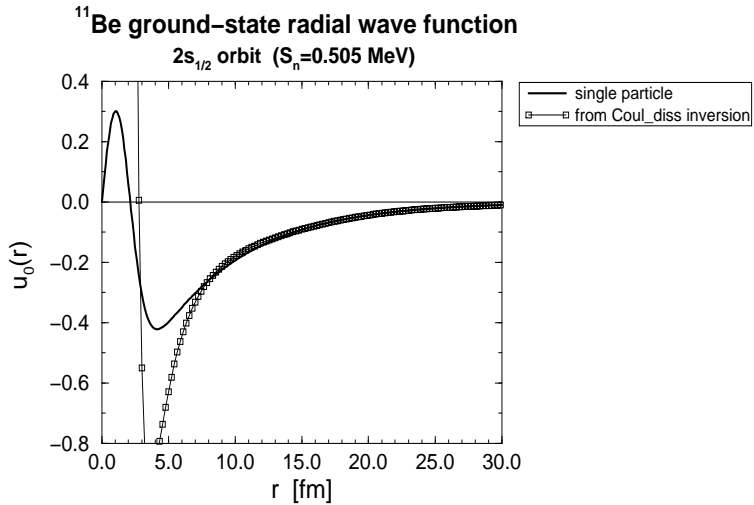}}
\end{center}
\caption{Reconstruction of the radial wave function of the
\mbox{ground-state} of ${}^{11}$Be. The \mbox{single-particle}
wave function calculated from a Woods-Saxon potential is shown
for comparison. See text for more explanations.
}
\label{fig:wf}
\end{figure}

We will now proceed to show how the bound state
wave function can be extracted from a measurement
of the E1 matrix elements. The principal ingredient
in the calculation of these is the overlap integral 
$ {\cal I}_{l_{b}l_{c}} $. As just shown, a 
plane-wave approximation for the continuum
wave function is a good approximation. Then,
the radial overlap becomes
\begin{equation}
{\cal I}_{01}(k) = 2i \sqrt{3 \pi} \int\limits_{0}^{\infty}
u_{0}(r) r^2 j_{1}(k r) dr
\label{eqi2}
\end{equation}
where $j_{1}(k r)$ is a spherical Bessel function.
${\cal I}_{01}(k)$ is nothing but the Fourier-Bessel
transform (or Hankel transform) of the
bound state wave-function $ u_{0}(r) $. It
can be promptly inverted to obtain
\begin{equation}
u_{0}(r) = \frac{1}{i \pi^{3/2}\sqrt{3}} \
\int\limits_{0}^{\infty}
{\cal I}_{01}(k) k^2 j_{1}(k r) dr .
\label{eqi3}
\end{equation}
This equation provides the radial component
of the bound state wave function in terms
of the radial matrix elements ${\cal I}_{01}(k)$
as a function of the neutron wave number $k$. 
These matrix elements can
be derived (see Eqs.~\ref{ME1},\ref{ME2})
from neutron capture measurements
as well as from a Coulomb dissociation
experiment.
%
%%%%%%%%%%%%%%%%%%%%%%%%%%%
%
\begin{figure}
\begin{center}
\leavevmode
\hbox{
\epsfxsize=10.0cm
\epsfysize=9.0cm
\epsffile{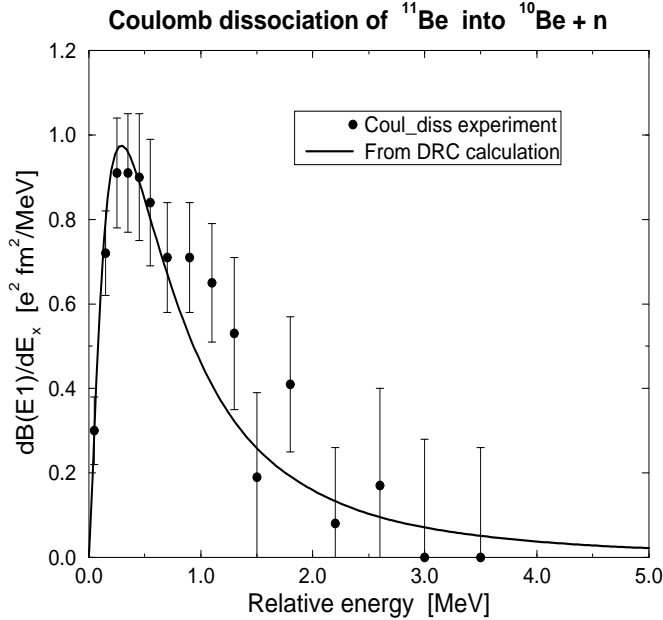}}
\end{center}
\caption{E1 strength distribution for the direct break-up
of ${}^{11}$Be into ${}^{10}$Be$ + n $ \protect\cite{Nax94}. 
}
\label{fig:11Be_diss}
\end{figure}
%
%%%%%%%%%%%%%%%%%%%%%%%%%%%
%
An example of the reconstruction of the radial part of a 
halo-state wave function is shown here. The $B(E1)$
strength distribution as measured in a Coulomb dissociation
experiment \cite{Nax94} is shown in figure \ref{fig:11Be_diss}. 
For comparison, we show in the figure a calculation made 
using a DRC model as described in the previous paragraphs. 
The solid-line of figure \ref{fig:11Be_diss} has been 
then utilized to obtain the radial part of the E1 
matrix elements as a function of the $ n + {}^{10}$Be 
relative energy.
Then, Eq.~\ref{eqi3} has been employed to calculate
the radial part of the ground-state wave function. The
result is shown in figure \ref{fig:wf}. The comparison is
made with the wave function obtained from a Woods-Saxon
potential with the depth adjusted to reproduce the
correct binding energy. A technique called 
\mbox{Pantis-integration} \cite{Pa75}
has been adopted to calculate the integral of Eq.~\ref{eqi3}.
We can conclude that the radial part of the halo-state
wave function could be reconstructed down to approximately
7 fm. 

In concluding we would like to stress here that this
technique is essentially {\it model-independent} as far
as the plane-wave approximation holds its validity
(and it does, for $l = $1 partial waves) in the
calculation of the continuum two-body wave function. 
The radial wave function of halo states can be therefore 
derived directly from neutron capture measurements and/or
from Coulomb-dissociation experiments. 
%
%%%%%%%%%%%%%%%%%%%%%%%%%%%
%
%
%%%%%%%%%%%%%%%%%%%%%%%%%%%
%

\section{Conclusion}
We have shown here the relations which can be established
between ``exotic'' nuclear structure properties (such as
neutron halo configurations) and the reaction mechanisms 
which lead to their formation. In particular we have shown
that spectroscopic information of halo
states can be derived from the knowledge of
E1 matrix elements for transitions 
between bound states and the continuum. 
These matrix elements can be derived from 
neutron capture measurements in the case of stable nuclei. 
For nuclei far from the stability line, we have shown that 
their structure can be investigated by Coulomb dissociation 
experiments. 
With the advent of machines capable of producing radioactive
nuclear beams, this technique already firmly established in experiments 
with light neutron-rich nuclei will represent a decisive tool
for the exploring the structure of intermediate mass and
heavy nuclei far-off 
the $\beta$-stability line.

\sectionnonum{Acknowledgments}
We would like to thank Prof.~Y.~Nagai and Dr.~T.~Shima of the Tokyo Institute
of Technology for useful discussions that greatly helped to
generate the present contribution.
This work has been partially supported by the Science and Technology
Agency through grant STA-194912.

\sectionnonum{References}

\end{document}